\documentclass{pasj00}
\newcommand{\inst}{\altaffilmark}

\draft

\begin{document}

\SetRunningHead{Yu--Peng Chen et al}{IGR J17098-3628 Observed by INTEGRAL and RXTE/ASM}
 
\title{INTEGRAL and RXTE/ASM Observations on IGR J17098-3628 }
 \author {Yu--Peng Chen \inst{1},Shu Zhang \inst{1},   Nick Schurch \inst{1,2}, Jian--Min Wang\inst{1,3}, Werner Collmar\inst{4},\\ Ti--Pei  Li \inst{1,5},  Jin--Lu Qu\inst{1},  Cheng--Min Zhang\inst{6}
}
   \altaffiltext{1} {Key Laboratory for Particle Astrophysics, Institute of High Energy Physics, 19B YuQuan Road, Beijing 100049, China}
\altaffiltext{2} {Department of Physics, University of Durham, South Road, Durham DH1 3LE, UK}
\altaffiltext{3} {Theoretical Physics Center for Science Facilities (TPCSF), CAS}
\altaffiltext{4} {Max-Planck-Institut f\"ur extraterrestrische Physik,
               P.O. Box 1603, D-85740 Garching, Germany  }
 \altaffiltext{5} {Center for Astrophysics,Tsinghua University, Beijing 100084, China}
\altaffiltext{6} {National Astronomical Observatories, Chinese Academic of Sciences, Beijing 100012, China}



\maketitle

\begin{abstract}
To probe further the possible nature of the unidentified source IGR J17098-3628, we have carried out
a detailed analysis of its long-term time variability as monitored
by RXTE/ASM, and of its hard X-ray properties as observed by
INTEGRAL.  INTEGRAL has monitored this sky
region over years and significantly detected  IGR J17098-3628 only
when the source was in this dubbed active state. In particular, at
$\ge$ 20 keV, IBIS/ISGRI caught an outburst in March 2005, lasting
for $\sim$5 days with detection significance  of 73$\sigma$ (20-40
keV) and with the emission at $< $200 keV. The ASM observations reveal that the soft X-ray lightcurve shows a similar outburst to that detected by INTEGRAL, however the peak of the soft X-ray lightcurve either lags, or is preceded by, the hard X-ray ($>$20 keV) outburst by $\sim$2 days. This resembles the behavior of X-ray novae like XN 1124-683, hence it further suggests  a LMXB nature for IGR J17098-3628. While the quality of the ASM data prevents us from drawing any definite conclusions, these discoveries are important clues that, coupled with future observations, will help to resolve the as yet unknown nature of IGR J17098-3628.

\end{abstract}

\section{Introduction}

INTEGRAL has detected roughly 500  sources at energies $\ge$
20 keV \citep{boda}. Among them, X-ray binary systems were identified for
32\% of the times, and another 26\% remained unidentified.
%
%
As one of the unidentified sources, IGR J17098-3628 was discovered
\citep{source} at the end of March 2005 by INTEGRAL/IBIS during the 
private Open Program observation dedicated to the deep view to the
Galactic center.
 The source was located at R.A.(2000) $=17^{h}09^{m}48^{s} $, Dec. $=-36\degree
 28\arcmin 12\arcsec $, with an error box of $2\arcmin$ at 90$\%$
confidence. The source got a peak flux at $\sim$ 60 and 95 mCrab for
18-45 and 45-80 keV, respectively. The   20-60 keV flux evolved from
$\sim50$ mCrab at 2005 March 26 00:00, to $\sim9$ mCrabs at April 3 23:30 (UTC) \citep{453}.
The preliminary spectral analysis \citep{spe} showed the source spectrum
became quite soft along this evolution. Following INTEGRAL, IGR J17098-3628 was observed by RXTE on March 29
of 2005, and was detected at 80 mCrab in the 3--20 keV band, with a
hard power-law tail of a spectral index $\sim$ 2.5. Then the source was assumed a
BHC and X-ray Nova (Grebenev et al. 2007). The column density was found to be less than 1$\times$10$^{22}$ atoms cm$^2$.
  
Later on, Swift observed this source on May 1 of 2005, and refined
 its location to R.A.\ (2000) $=17^{\rm h}09^{\rm m}45.9^{\rm s} $, Dec.\
$=-36\degree 27\arcmin 57\arcsec $, with 90$\%$ confidence, and an
uncertain radius of 5 arcseconds \citep{kenn}. This made the search
possible for the potential counterparts in optical and  infrared
bands. From the 2MASS ALL-SKY Catalog a possible counterpart was found,
J17094612-3627573 \citep{kong}. However, the most recent optical and
infrared observations, made with the 6.5m Magellan-Baade telescope,
revealed that 2MASS J17094612-3627573 is in fact composed off
several sources \citep{stee}. Furthermore, Very Large Array (VLA)
observations of IGR J17098-3628 were made on March 31, April 5, and
May 4, 2005 at 4.86 GHz \citep{radio}, which  led to the discovery of a
radio transient at 0.8 arcseconds from the Swift XRT position, and
hence regarded as the possible radio counterpart. Finally on July 9,
Swift/XRT observed IGR J17098-3628 again and found that the spectrum was  fitted by an
absorbed disk blackbody model. The column density obtained was
(0.89$\pm$0.02$)\times$10$^{22}$ atoms cm$^{-2}$ \citep{kenn2}.

In this paper, we report the analysis of all available observations
on IGR J17098-3628  carried out by IBIS/ISGRI and JEMX onboard
INTEGRAL, and the All Sky Monitor (ASM) onboard RXTE. This allows us
to trace the source behavior in X-rays back to 1996 with ASM and
back to 2002 with INTEGRAL. We put the outburst in 2005 in this context.

\section{Time variability monitored by  ASM}

ASM is one of the three detectors onboard the RXTE
satellite \citep{gruber96,swank94}, which  has been used to track the
long-term behavior of the source in the energy band 1.5-12 keV since
February 1996. The target source was usually observed several times
per day, within the so-called dwells, of duration 96 seconds each.
The extracted source light curves are presented for the energy bands of 1.5--3, 3--5, 5--12, and
1.5--12 keV, respectively.

\begin{figure}[ptbptbptb]
\centering
 \includegraphics[angle=0, scale=0.7]{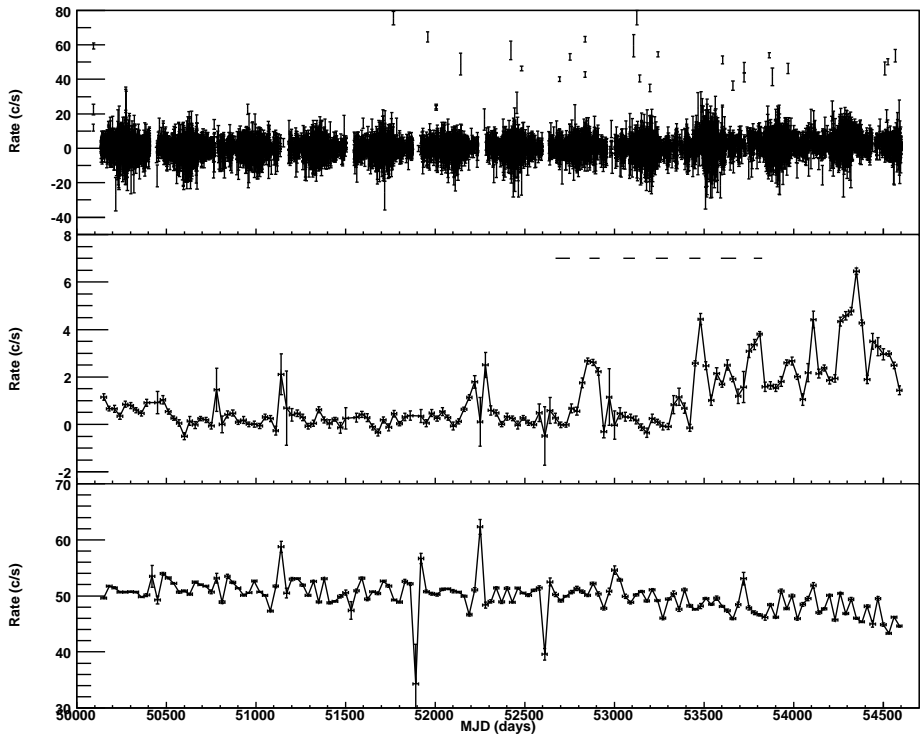}
      \caption{ASM light curves (1.5--12 keV) of IGR J17098-3628 with each bin representing a dwell (top) or average over month  (middle) timescales
      between 1996 and 2008. In the middle panel, the available INTEGRAL observations are shown by the bars. ASM light curve (1.5--12 keV) of GX349+2
      is presented in the bottom panel, with each bin representing average over month timescales  between 1996 and 2008. }
         \label{asm_lc}
\end{figure}

IGR J17098-3628 has been monitored by ASM  as one
of the most active sources ever seen since 2002.  As shown in
 Figure \ref{asm_lc}, the 1.5--12 keV lightcurves,  retrieved
in a time period from January 1996 to May 2008, show the
variability, where each bin represents a dwell, or --in the middle
panel-- the average over month  timescale. Many individual
bursts-like events existing in the dwell lightcurve can be spotted, and the flux
maxima can sometimes reach 65.5 cts/s (the dwell at
MJD=51958.219410), corresponding to 873 mCrab in the 1.5-12 keV band.
We have made communication with the ASM team for consulting on such feature.
 It turns out that the source IGR J17098-3628 is only 1 degree away from GX349+2, a very strong
X-ray source.  The angular resolution of each of the ASM cameras is
much better in one coordinate $\phi$ ($\sim$3$\arcmin$) than in the other coordinate
$\theta$ ($\sim$15$\arcmin$).    This is the kind of problem that is (understandably) difficult for the ASM
analysis system to properly handle; similar problems have arisen with
other sources. But in general, GX349+2 is stable in the 1.5-12 keV band observed by ASM (see Figure \ref{asm_lc}, the bottom panel).
In short, the analysis software has incorrectly
attributed the flux from GX349+2 to IGR J17098-3628 only on those
relatively few occasions. Actually, 
the big uncertainty in this lightcurve is the contamination from the neighboring source, IGR J17091-3624, located (only 
$0.17 \degree$ away) too close to IGR J17098-3628 to be resolved by ASM. As the result, the burst events on monthly timescale 
as shown in the middle panel of Fig.1 may 
sometimes attribute to  IGR J17091-3624. Fortunately, this problem can be well solved via INTEGRAL's mapping this region with 
either ISGRI or JEMX. In the monthly lightcurve there are several well established
outbursts with a duration of several months.  The first one occurs
around MJD 52800 (June, 2003), with averaged peak flux reaching 40 mCrab level. We will see later that, this outburst came 
from the IGR J17091-3624 (Grebenev et al. 2007), and the ones thereafter observed by INTEGRAL were from  IGR J17098-3628. 
IGR J17091-3624 seems to be in quiet most of time since  MJD 52800 (Grebenev et al. 2007).  
A sequence of outbursts have happened since MJD 53455 (March, 2005), and the
1.5--12 keV flux has kept staying at the flux level of $\sim$ 50
mCrab over years. These results, accompanying with the INTEGRAl results which will be shown later on, 
suggest IGR J17098-3628 is  in  active phase within the last several years.

\section{Hard X-ray properties revealed by INTEGRAL}
\subsection{Observations and data analysis}

INTEGRAL is an ESA scientific mission (Winkler et
al.  2003) dedicated to spectroscopy
($E/\Delta E$$\simeq$500; SPI see Vedrenne et al. 2003) and imaging (angular
resolution: 12$\arcmin$ FWHM, source location accuracy:
$\simeq$1--3$\arcmin$; IBIS, see Ubertini et al. 2003) of celestial
$\gamma$-ray sources in the energy range 15\,keV to 10\,MeV, with
simultaneous monitoring in the X-ray (3--35\,keV, angular
resolution: 3$\arcmin$; JEMX, see Lund et al. 2003)  and optical (V-band,
550\,nm; OMC, see Mas-Hesse et al. 2003)
 energy ranges.
All the instruments onboard  INTEGRAL, except the OMC, work with
coded masks.
The observational data from the detectors IBIS/ISGRI (15-200 keV) and JEMX
have been adopted in our analysis of IGR J17098-3628.

The available INTEGRAL observations when IGR J17098-3628 falls into
the Fully Coded Field of View (FCFoV) of ISGRI comprise about 2400
science windows (scw's, typically 2000 seconds each), adding to a
total exposure of 3700 ks. Most of these observations are  carried
out in a 5x5 dithering mode. The data are subdivided into 9 groups
according to the sequence in observational time. The groups T5, T6,
and T7 represent the observations made one month  prior to the hard
X-ray outburst, 6 days in 2005 March 23-28 --during the outburst,  and one
month after the outburst, respectively (see Table 1 for a summary of
the individual observational groups).

The analysis of JEMX and ISGRI data is performed by using the
INTEGRAL Offline Scientific Analysis (OSA) software, version 6.0.
All the sources within the FoV that  are brighter than or comparable
to IGR J17098-3628  are taken into account in extracting the source
spectrum.   An additional  2\% systematic error are added to the
spectra because of calibration uncertainties. The ISGRI spectrum was normalized to the JEMX spectrum by a factor $0.77^{+0.07}_{-0.06}$ derived from the fit. The
spectra are fitted with XSPEC v12.3.1 and the model parameters are
estimated  with 90$\%$ confidence level.
\subsection{INTEGRAL results}
\subsubsection{Sky maps }
We have looked into the skymaps of both ISGRI and JEMX to see for each of the 9 time zones which source was on active. The detection 
significances from ISGRI in the 20-40 keV band are summarised in Table 1. We find that
IGR J17091-3624 was in active phase within  2003 April and 2004 April (T1--T3), and  IGR J17098-3628 took the role since 2005 March 
in T6. After 2004 August (in T4), IGR J17091-3624  was not detectable by INTEGRAL at the X-rays.

\begin{figure}[ptbptbptb]
     \centering
     \includegraphics[angle=0, scale=0.7]{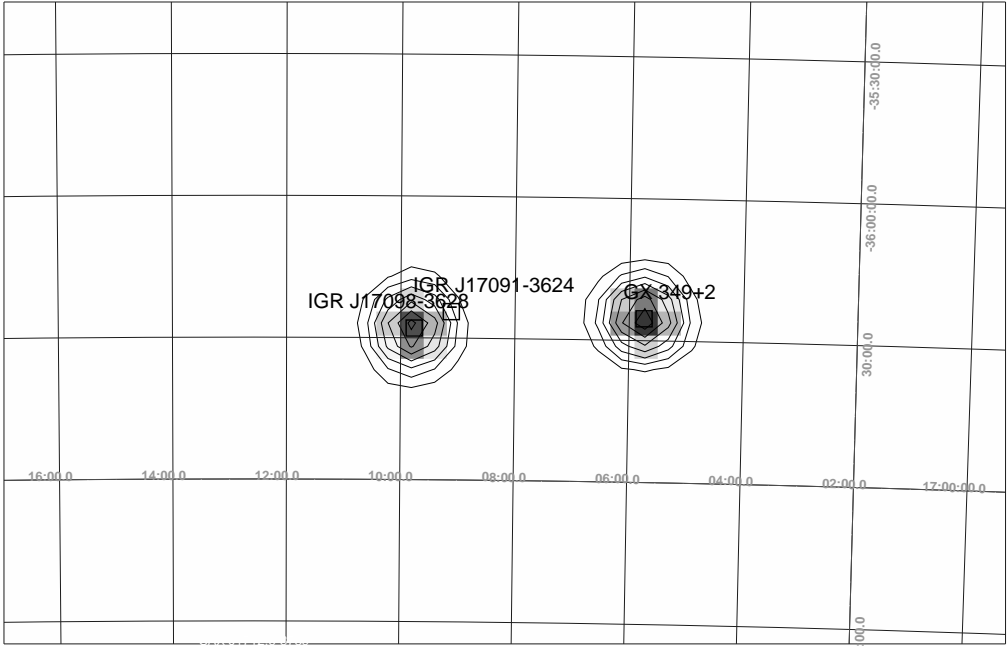}
     \caption{The 20-40 keV significance map from IBIS/ISGRI of IGR J17091-3624. The contours start at detection significance level of 10 $\sigma$, with the step of 10 $\sigma$ for the time period during the hard X-ray outburst (March 23--28, 2005).
   }
     \label{skymap2}
  \end{figure}

\begin{figure}[ptbptbptb]
     \centering
     \includegraphics[angle=0, scale=0.7]{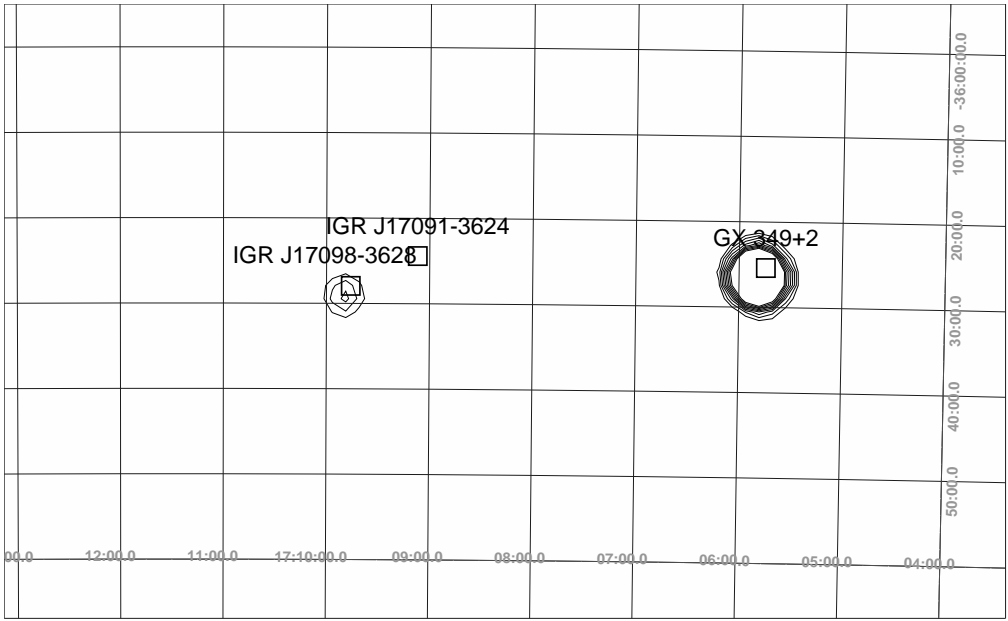}
     \caption{The 3-6 keV significance map from JEMX of IGR J17098-3628. The contours start at detection significance level of 10 $\sigma$, with the step of 5 $\sigma$ for the time period during April 2--21, 2005.}
     \label{skymap1}
  \end{figure}

IGR J17098-3628 was detected by ISGRI at energies $\ge$ 20 keV
during March 24-28, 2005 when the source had an outburst in hard X-rays.
The detection significances were 73.4$\sigma$ in the 20--40 keV band
(for the mosaic image seen in Figure \ref{skymap2}), 38$\sigma$ in
the 40--60 keV band, 21 $\sigma$ in the 60--100 keV band, and
8$\sigma$ in the 100--200 keV band. The source was  still visible
during its decay in the following one month (T7), with the
detections significances dropping to 16.9$\sigma$ in 20--40 keV. For
all other INTEGRAL observations, only marginal detections or upper
limits were obtained at energies $\ge$ 20 keV, e.g., for the time
period  T9, the source was detected only in 20-40 keV with
significance $\sim$ 8$\sigma$ level (see Table 1). Since JEMX overlaps with ASM
in most of its working energies, the detections at $\le$ 15 keV
follow in general the same trend indicated by the
 ASM light curves. Figure \ref{skymap1} shows the most significant JEMX
 detection of the source in March 31-April 21, 2005, which gives $\sim$
22 $\sigma$ in the 3--6 keV band.

\subsubsection{Light curves }

\begin{figure}[ptbptbptb]

  \includegraphics[angle=0, scale=0.7] {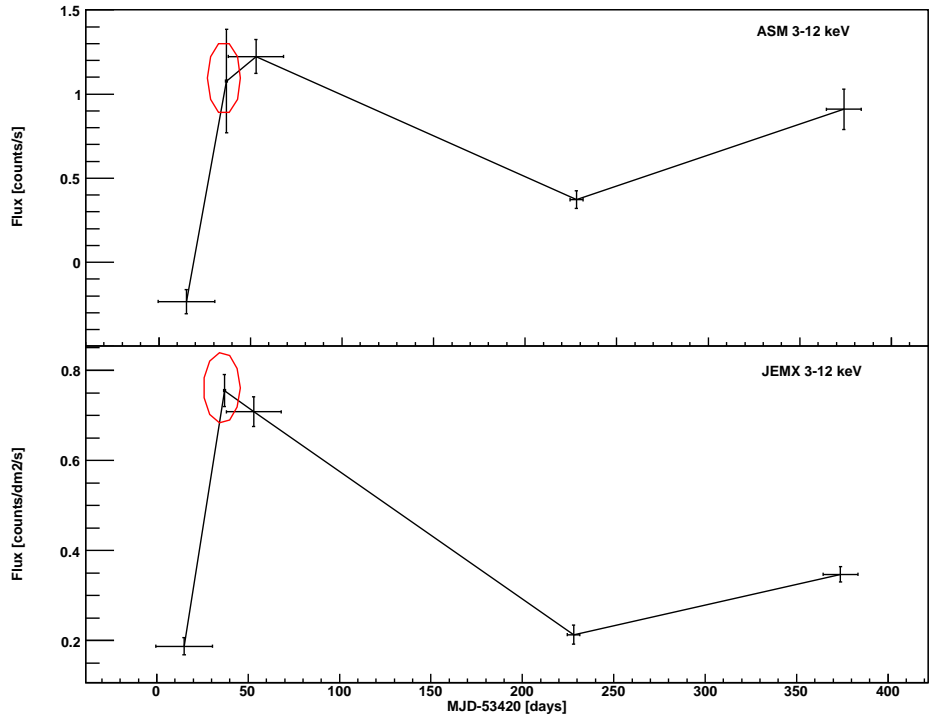}
     \caption{Lightcurves from ASM and JEMX in the 3--12 keV band, with each bin averaged over the individual data group in T5-T9. The hard X-ray outburst is marked by the red circle.  }
         \label{asm-jemx}
   \end{figure}

\begin{figure}[ptbptbptb]
\includegraphics[angle=0, scale=0.7] {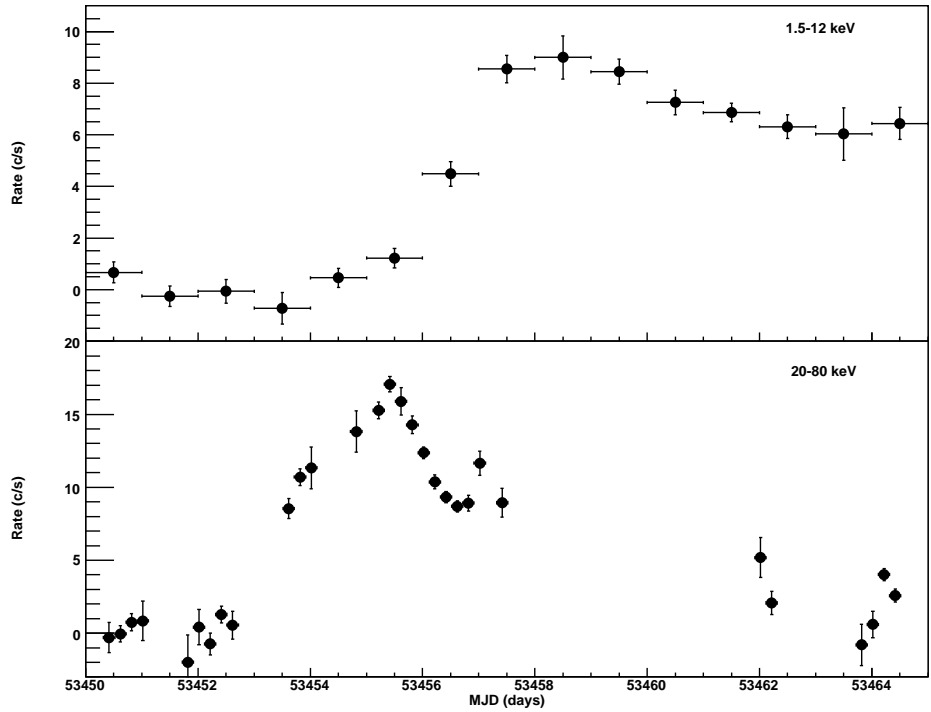}
\caption{Lightcurves of IGR J17098-3628 covering the March 2005 days-duration outburst. {\it Upper panel}: 1.5-12 keV ASM lightcurve (1-day binsize). {\it Lower Panel}: 20-80 keV INTEGRAL/ISGRI lightcurve (0.2-day bin size).}
\label{asm-integral}
\end{figure}

\begin{figure}[ptbptbptb]

  \includegraphics[angle=0, scale=0.7] {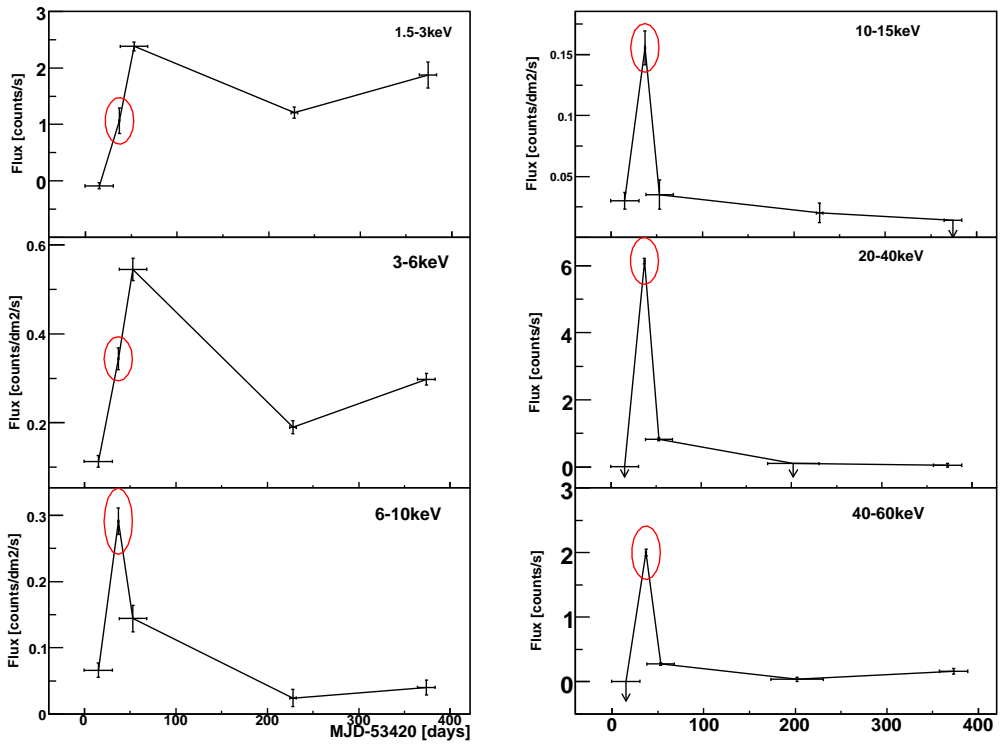}
     \caption{Lightcurves from ASM (1.5--3 keV), JEMX (3--6keV, 6--10, 10--15 keV) and ISGRI (20--40 keV, 40--60keV), with each bin averaged over the individual data group T5-T9. The hard X-ray outburst is marked by the red circle.  }
         \label{asm-jemx-isgri}
   \end{figure}

We use version 6.0 of the INTEGRAL Offline Scientific Analysis (OSA) software to construct the $>$20 keV ISGRI X-ray 
lightcurve of this outburst and compare it to the 1.5-12 keV ASM lightcurve from the same period. 
Unfortunately IGR J17098-3628 was either prohibitively faint or out of the FoV for JEMX for most of the INTEGRAL
observation, and the remaining data are contaminated by a strong source in the JEMX field-of-view
(this is a known issue for JEMX data extraction - INTEGRAL Help Desk Priv. Comm.) to prevent from producing the proper
lightcurve through running the pipeline. We therefore generate the JEMX lightcurves via reading out the flux 
at the source position from the mosaic map. A comparison of the JEMX and ASM lightcurves at 3-12 keV is shown in Figure \ref{asm-jemx}, where
the trends are consistent with each other. We find that the soft X-ray lightcurve shows a similar outburst to that detected by INTEGRAL, 
however the peak of the soft X-ray lightcurve lags the hard X-rays by $\sim$2 days (Figure \ref{asm-integral}).
We note, however, that no hard X-ray data cover the peak outburst observed in the ASM data. 
If there is a hard X-ray peak that coincides with the soft X-ray peak, then, since there is no evidence for two peaks 
in the ASM data, the earlier hard X-ray peak may well be interpreted a precursor event to a wider band concurrent X-ray outburst rather than evidence for a lag between the hard and soft X-ray bands.  Figure \ref{asm-jemx-isgri} shows a comparison of the flux variability in
the time zones T5--T9, as obtained with ASM in 1.5--3 keV, JEMX in 3--6,
6--10, 10--15 keV, and ISGRI in 20--40 keV, and 40--60 keV bands.

\subsubsection{Spectra}

\begin{figure}[ptbptbptb]
  \includegraphics[angle=-90, scale=0.7] {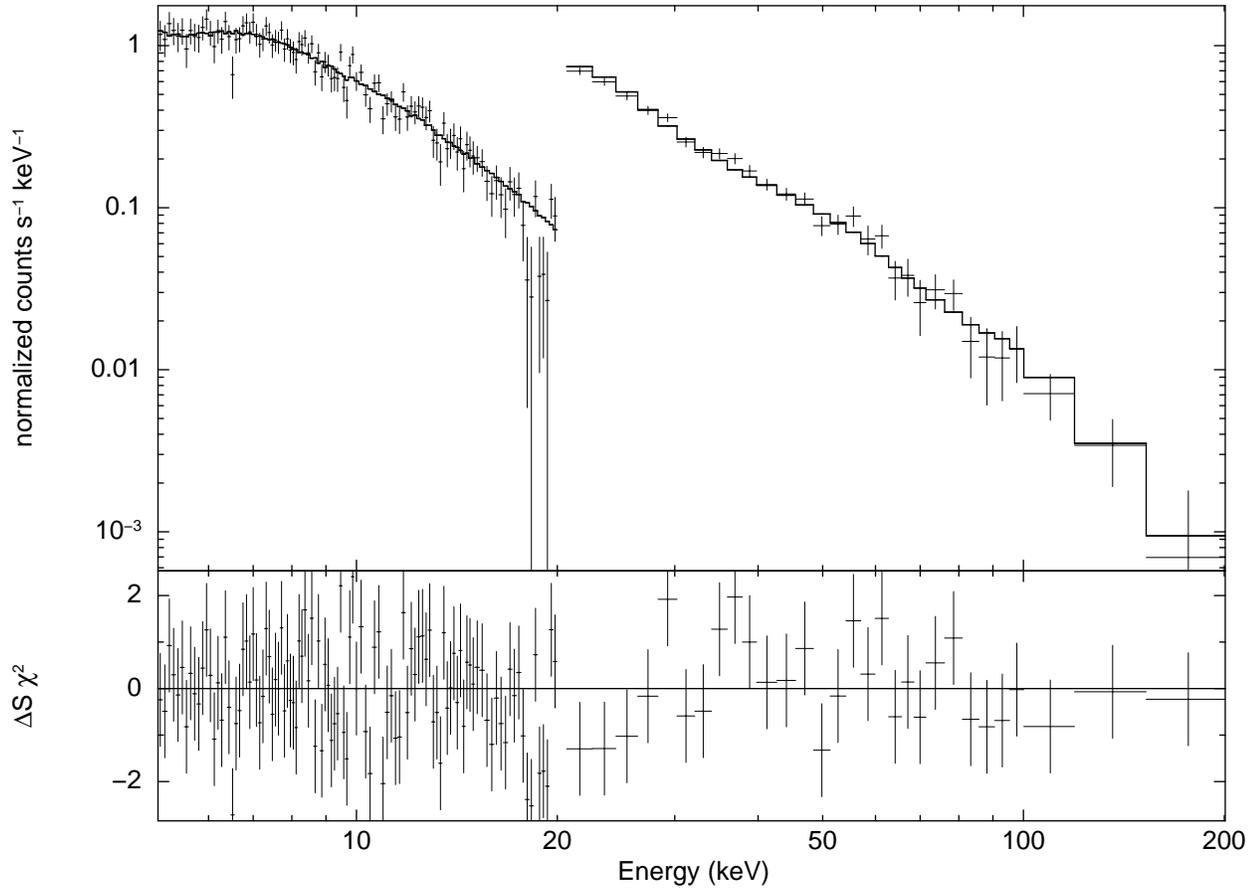}
     \caption{Spectral fit of a broken power-law model to the combined ISGRI/JEMX data obtained during the decaying phase of the hard X-ray outburst. }
         \label{spe_3}
   \end{figure}

Since the lightcurves show quite different behavior between the higher and lower energy bands,
it would be interesting to investigate the spectrum at the different periods of the outburst,
i.e., the time evolution of the spectrum along the outburst, although the overall spectra have been presented in Grebenev et al (2007) from using ISGRI data of
the outburst, supplemented by the available PCA/RXTE data.
For investigating the spectrum of the source, the time zone where the outburst was recorded
 by INTEGRAL (T6) was subdivided
into three smaller time periods according to observations:
March 23st 03:13 to 25st 02:35(the rising phase), 25st 23:03 to 27st 00:51 ( the peaking phase) and 27st 00:53 to
28st 11:15 (UTC) (the decaying phase).  Due to the relatively small
FoV of the JEMX, its data are only available during the decaying phase,
which is then combined with the corresponding ISGRI data
for a broader spectral fitting.
 Firstly we have
tried different fitting models in the 20--200 keV band only using the ISGRI data: simple power-law, power-law with cutoff, and
broken power-law shapes. The goodness of each trial
is shown in Table 2. We find that the relatively
poor statistics of the data from
 the rising of the hard X-ray outburst  prevents from discriminating
the models; the
power-law with cutoff model might be the best choice when the outburst
reached the maximum; and the
power-law model shows a good fit to the data from the decay of the
outburst onwards. From the fit of a simple power law model, the spectral index changes from
$\sim$ 1.7 when rising to roughly 2.6 when decaying. We tried as well a disk blackbody model for the decaying phase (the broadband 5-200 keV spectrum), and found
the reduced $\chi^2$ ($\sim$1.36) gets worse than that from the best-fit model of a broken power law (see Figure \ref{spe_3}).
The best-fit parameters of
the proper model are then presented in Tables 3, 4, and 5, for the three phases, respectively. 
That the different episodes of the
outburst have to be described by different models might suggest rather
strong time evolution.   By taking the
distance of 10.5 kpc  as estimated by Grebenev et al.(2007), the source had a
luminosity of $\sim$  $9.8\times 10^{36}$erg s$^{-1}$ averaged over
the whole outburst in the 20--200 keV. The luminosities in individual phase are  $5.3\times 10^{36}$erg s$^{-1}$ (the rising), 
$1.6\times 10^{37}$erg s$^{-1}$ (the peaking) and $8.9\times 10^{36}$erg s$^{-1}$ (the decaying).

\section{Discussion and summary}

IGR J17098-3628 may have stepped into an active phase in the beginning of 2005. INTEGRAL detected a days-long 
outburst extending up to energies of $\sim$200 keV. A similar outburst is detected in the soft X-ray (1.5-12 keV) band, 
however the onset of the outburst lags the hard X-ray outburst by $\sim$2 days, in a fashion reminiscent of X-ray novae 
like XN 1124-683 \citep{nova}.
The spectral analysis of the initial stages of the outburst detected by INTEGRAL supports the association with X-ray novae 
(Grebenev et al. 2007), the cool disk temperature and small inner radius of the accretion disk both suggest that 
IGR J17098-3628 is a newly discovered black hole system.  Current theoretical models of the formation of X-ray novae 
include the mass transfer instability (MTI) model \citep{hkl86}, the disk thermal instability  (DTI) model
(Cannizzo et al. 1982; Faulkner et al. 1983; Meyer \& Meyer-Hofmeister 1984; Huang \& Wheeler 1989; Mineshige \& Wheeler 1989; Ichikawa et al. 1994)
 Unfortunately, the only way to discriminate between these models relies
on detailed differences in the observed soft energy spectrum of the source and so the lack of any detailed X-ray spectra 
of the IGR J17098-3628 outburst in this energy band  prevents us from drawing any conclusions on the X-ray nova mechanism 
for IGR J17098-3628. That the three episodes of the hard X-ray outburst observed by INTEGRAL have to be described by different 
models might suggest rather strong time evolution. i.e.,   the spectral index changed from $\sim$ 1.7 when rising to 
roughly 2.6 when decaying suggesting the cooling of the source.

Although the time variability of X-ray novae is not well understood theoretically, observations of a variety of 
X-ray novae not only show that the hard X-ray flux often rises and peaks earlier than the outburst in soft X-rays, 
but can also show precursor events at hard X-rays several days prior to the main outburst (Chen et al. 1997). 
While it is not possible to associate the outburst in IGR J17098-3628 with any particular X-ray nova mechanism spectrally, 
it is clear that the typical accretion scenario for the low mass X-ray binaries holding a black hole is not capable of reproducing the observed lag 
between the hard and soft X-ray outbursts. In this scenario, the soft X-rays are produced by the accretion disk, 
while the hard X-rays are produced by comptonization of soft seed photons in the inner part of the accretion disk 
\citep{disalvo,stella}. Clearly, large changes
in the hard X-ray flux can, at best, be coeval with large changes in the soft X-ray seed photons and cannot precede
variations of the soft X-ray photons that the process of comptonization relies on to produce the hard X-ray tail.
One potential mechanism that may reproduce the observed lag is for an instability in the disk to trigger the formation 
of a temporary jet, which produces hard X-rays via inverse comptonization, in advance of any change in the flux of soft 
X-rays from the disk.

A more straightforward scenario in understanding this different time behaviour  between the hard and the soft X-rays
may lie in the tie-up with  the structure of the accretion flow.
The precursor event can be explained as the transition from the "hard" state to "soft" state in the BHC system.
The accretion time in the inner parts of the disk is much smaller
than the duration of X-ray nova's outburst therefore changes in the
spectral state of these sources are connected with general changes in
structure of the accretion flow rather than in changes in number of
soft seed photons for Comptonization. Observations show that all BH
systems have the hard Comptonized spectrum at low
accretion rates and the soft DBB(disc blackbody) spectrum (with a possible weak hard
tail) at high accretion rates. This may be connected with approaching
of the inner radius of a cold accretion disk to the radius of
marginally stable orbit when the accretion rate increases. The
observed lag between outbursts in hard X-rays and soft X-ray may be
directly connected with this observed dependence: the accretion rate
is small at the initial stage of the outburst thus the spectrum is
hard and the outburst is observed in hard X-rays, but later the
accretion rate rises and the spectrum becomes soft thus the outburst
is observed in soft X-rays.

Although no clear statement can be made regarding their nature, the discovery  the lag/precursor event revealed by the concurrent RXTE ASM and INTEGRAL observations in 2005, is important piece of the IGR J17098-3628 puzzle that will help to resolve the as yet unknown nature of this perplexing source.

\section{Acknowledgement}
  The authors are highly grateful to the anonymous referee for his/her many valuable comments that are great helpful to improve the paper. The authors would like to thank Prof. Cui Wei for his many valuable comments, especially regarding the details of the ASM data. This work was subsidised by the National Natural Science Foundation of China, and the CAS key Project KJCX2-YW-T03. J.-M. Wang would like to thank the Natural Science Foundation of China for support via NSFC-10325313, 10521001, 10733010, C.-M. Zhang would like to thank the Natural Science Foundation of China for support via NSFC-10773017, and N. J. Schurch would like to thank the Department for Innovation, Universities and Skills (UK) for support via a UK-China Fellowship for Excellence.

\newpage
\begin{table}[ptbptbptb]
\begin{center}
\label{tab1}
\caption{INTEGRAL IBIS/ISGRI Observations of IGR J17091-3624 and J17098-3628 in the 20--40 keV band for those with the detection significance (signal to noise ratio) larger than 3$\sigma$.}
\vspace{5pt}
\begin{tabular}{cccccccccc}
\hline \hline
 Label & Revolution & Start Date  & End Date &  Exposure (ks)  & IGR J17091-3624 & IGR J17098-3628\\
       &            &             &          &              & counts/s (significance)& counts/s (significance)\\
\hline
 T1 &0037--0063 &  2003-02-01  &  2003-04-22  & 593 &  0.35(8.6) & -\\
\hline
 T2 & 0100--0120 &  2003-08-09 &  2003-10-07  & 652 &1.64(41.8) & -\\
\hline
 T3 &0164--0185 &  2004-02-16  &  2004-04-19  & 360 & 2.51(44.9) & -\\
\hline
 T4 &0224--0246 &  2004-08-15  &  2004-10-20  & 477 & - & -\\
\hline
T5 & 0286--0297 &  2005-02-15  &  2005-03-22  & 456 &  - & -\\
\hline
T6  &0298--0299 &  2005-03-23  &  2005-03-28  & 152 &  - & 6.08(73.4)\\
\hline
T7  &0300--0307 &  2005-04-02  &  2005-04-21  & 341 &  - & 0.96(16.9)\\
\hline
T8  & 0345--0370 &  2005-08-10  &  2005-10-26  & 386 & - & -\\
\hline
T9  &0406--0423 &  2006-02-09  &  2006-04-03  & 337 &  - & 0.49(7.8)\\
\hline
\end{tabular}
\end{center}
\end{table}

\newpage

\begin{table}[ptbptbptb]
\begin{center}
\label{tab2}
\caption{The spectral index and reduced $\chi^2$ resulting from the fitting of
ISGRI data in the 20--200 keV band with different models}
\vspace{5pt}
\begin{tabular}{ccccccc}
\hline
\hline
Date &  Pow law & Cutoff pow law  &   Broken pow law \\
(phase)    &  $  \alpha$ &  $  \alpha$ &  $  \alpha$ \\
    & ($\chi^2$/$\nu$) & ($\chi^2$/$\nu$) & ($\chi^2$/$\nu$) \\
\hline
03-23 03:13..03-25 02:35 & $1.74^{+0.27}_{-0.25}$ & $1.61^{+0.3}_{-1.8}(<464 keV)$ & $1.74^{+0.25}_{-0.25}(>3.7 keV)$ \\
 (rising phase) & (0.71) & (0.78) & (0.88) \\
\hline
03-25 23:03..03-27 00:51 & $2.13^{+0.06}_{-0.06}$ & $1.54^{+0.27}_{-0.27}(<82 keV)$ & $2.13^{+0.06}_{-0.06}(>5.5 keV)$ \\
  (peaking phase) & (1.49) & (0.95) & (1.60) \\
\hline
03-27 00:53..03-28 11:15 & $2.60^{+0.08}_{-0.08}$ & $2.16^{+0.33}_{-0.36}(<95 keV)$ & $2.60^{+0.08}_{-0.08}(>7.4 keV)$ \\
  (decaying phase) & (0.86) & (0.72) & (0.93) \\
\hline
\end{tabular}
\end{center}
\end{table}

\newpage

\begin{table}[ptbptbptb]
\begin{center}
\caption{Modelling to the ISGRI data from the rising phase of the hard X-ray outburst.}
\vspace{5pt}
\label{fit_1}
\begin{tabular}{ccccccc}
\hline
\hline
model & $N$& $  \alpha$&  $\chi^2$/$\nu$&   d.o.f &   $F$(20--200 keV)\\
   & ${\rm keV}^{-1}\,{\rm cm}^{-2}\,{\rm s}^{-1}$ at 1 keV  &   &   &   &    erg cm$^{-2}$ s$^{-1}$ \\
\hline
  powerlaw &  $0.036^{+0.064}_{-0.036}$& $1.74^{+0.27}_{-0.25}$& 0.71& 10& $4.0\times 10^{-10}$\\
\hline
\end{tabular}
\end{center}
\end{table}

\newpage

\begin{table}[ptbptbptb]
\begin{center}
\caption{Modelling to the ISGRI data from the peaking phase of the hard X-ray outburst.}
\vspace{5pt}
\label{fit_2}
\begin{tabular}{ccccccc}
\hline
\hline
model & $N$& $  E_{\kappa}$&$  \alpha$&  $\chi^2$/$\nu$&   d.o.f &   $F$(20-200 keV)\\
   & ${\rm keV}^{-1}\,{\rm cm}^{-2}\,{\rm s}^{-1}$ at 1 keV  &   keV &   &  &   &    erg cm$^{-2}$ s$^{-1}$ \\
\hline
  cutoffpl &  $0.12^{+0.14}_{-0.07}$& $82^{+65}_{-26}$& $1.54^{+0.27}_{-0.27}$& 0.95& 27& $1.2\times 10^{-09}$\\
\hline
\end{tabular}
\end{center}
\end{table}

\newpage

\begin{table}[ptbptbptb]
\begin{center}
\caption{Modelling to the combined ISGRI/JEMX data from the decaying phase of the hard X-ray outburst.}
\vspace{5pt}
\label{fit_3}
\begin{tabular}{cccccccc}
\hline
\hline
model & $N$&$  \alpha_{1}$& $  E_{break}$&$  \alpha_{2}$&  $\chi^2$/$\nu$&   d.o.f &   $F$(5-200 keV)\\
   & ${\rm keV}^{-1}\,{\rm cm}^{-2}\,{\rm s}^{-1}$ at 1 keV  &  &   keV &   &  &   &    erg cm$^{-2}$ s$^{-1}$ \\
\hline
  bknpower &  $0.08^{+0.14}_{-0.06}$& $0.85^{+0.53}_{-0.79}$& $7.19^{+0.28}_{-0.49}$& $2.66^{+0.07}_{-0.07}$&1.07& 133& $1.9\times 10^{-09}$\\
\hline
\end{tabular}
\end{center}
\end{table}

\newpage

\newpage

\begin{thebibliography}{}

\bibitem[Bodaghee et al. (2007) ]{boda}
    Bodaghee, A., Courvoisier, T. J.-L., Rodriguez, J., et al. 2007, A\&A, 467, 585
\bibitem[Cannizzo et al. (1982) ]{cgw82}
 Cannizzo, J.K., Ghosh, P., \& Wheeler, J.C.\ 1982, ApJ, 260, L83
\bibitem[Chen et al. (1997) ]{nova}
  Chen, W. , Shrader, C.R., \& Mario L. 1997, ApJ, 491, 312
\bibitem[Di Salvo et al. (2004) ]{disalvo}
  Di Salvo, T., Santangelo, A., \& Segreto, A. 2004, NuPhS, 132, 446
\bibitem[Di Salvo, \& Stella (2002) ]{stella}
  Di Salvo, T., \& Stella, L. 2002, to appear in the XXII Moriond Astrophysics
  Meeting "The Gamma-Ray Universe" (Les Arcs, March 9-16, 2002), eds. A. Goldwurm,
  D. Neumann, and J. Tran Thanh Van, The Gioi Publishers (Vietnam), astro-ph/0207219
\bibitem[Faulkner et al. (1983) ]{flp83} Faulkner, J., Lin, D.N.C., \& Papaloizou, J.\ 1983, MNRAS, 205, 359
\bibitem[Grebenev et al. (2005a) ]{source}
  Grebenev, S.A., Molkov, S.V., \& Sunyaev R.A.\ 2005a, ATel, 444
\bibitem[Grebenev et al. (2005b) ]{spe}
  Grebenev, S.A., Molkov, S.V., Revnivtsev, M.G., \& Sunyaev, R.A.\ 2005b, ATel, 447
\bibitem[Grebenev et al. (2007) ]{gre07}
  Grebenev, S. A., Molcov, S. V., Revnivtsev, M.G., \& Sunyaev, R.A.\ 2007, Proceedings of the 6th INTEGRAL Workshop 'The Obscured Universe',  ESA SP-622, page 373, astro-ph/07092313
\bibitem[Gruber et al. (1996) ]{gruber96}
  Gruber, D.E., Blanco, P.R., Heindl, W.A., et al. 1996, A\&AS, 120, 641
 \bibitem[Hameury et al. (1986) ]{hkl86} Hameury, J.M., King, A.R., \& Lasota, J.-P.\ 1986, A\&A, 162, 71
\bibitem[Huang \& Wheeler (1989) ]{hw89} Huang, M., \& Wheeler, J.C.\ 1989, ApJ, 343, 229
\bibitem[Ichikawa et al. (1994) ]{ich94} Ichikawa, S., Mineshige, S., \& Kato, T.\  1994, ApJ, 435, 748
\bibitem[Kennea  et al. (2005) ]{kenn}
  Kennea, J.A., Burrows, D.N., Nousek, J.A., et al. 2005, ATel, 476
\bibitem[Kennea et al.  (2007) ]{kenn2}
     Kennea, J.A., Capitanio, F. 2007, ATel, 1140
\bibitem[Kong (2005) ]{kong}
     Kong, A.K.H. 2005, ATel, 477
\bibitem[Lund et al. (2003) ]{jemx}
  Lund, N., Budtz-J\o rgensen, C.,  Westergaard, N.J., et al. 2003, A\&A, 411, L231
\bibitem[Mas-Hesse et al. (2003) ]{omc}
  Mas-Hesse, J.M., Gim\'enez, A., Culhane, J.L., et al. 2003, A\&A, 411, L261
\bibitem[Meyer \& Meyer-Hofmeister (1984) ]{mmh84} Meyer, F., \& Meyer-Hofmeister, E.\ 1984, A\&A, 132, 143
\bibitem[Mineshige \& Wheeler  (1989) ]{mw89} Mineshige, S., \& Wheeler, J.C. \ 1989, ApJ, 343, 241
\bibitem[Mowlavi et al. (2005) ]{453}
  Mowlavi, N., Kuulkers, E., Rodriguez, J., et al. 2005, ATel, 453
\bibitem[Rupen et al. (2005) ]{radio}
 Rupen, M.P., Mioduszewski, A.J., \& Dhawan, V. 2005, ATel, 490
\bibitem[Steeghs et al. (2005) ]{stee}
   Steeghs, D., Torres, M.A.P., Jonker, P.G., et al. 2005, ATel, 478
 \bibitem[Swank  (1994) ]{swank94}
  Swank, J.H. 1994, AAS, 185.6701
\bibitem[Ubertini et al. (2003) ]{ibis}
  Ubertini, P., Lebrun, F., Di Cocco, G., et al. 2003, A\&A, 411, L131
 \bibitem[Vedrenne et al. (2003) ]{esa}
  Vedrenne, G., Roques, J.-P., Sch\"onfelder, V., et al. 2003, A\&A, 411, L63
 \bibitem[Winkler et al. (2003) ]{Win}
   Winkler, C. Courvoisier, T. J.-L., Di Cocco, G., et al. 2003, A\&A, 411, L1


\end{thebibliography}
\end{document}